\documentclass[9pt,twocolumn,twoside]{opticajnl}
\journal{opticajournal} 
\doi{} 

\setboolean{shortarticle}{true}

\usepackage{graphicx}
\usepackage{booktabs}
\usepackage{tabularx}

\usepackage{lineno}

\title{Modular Programmable Photonics with Unitary Tiles}
\author[]{Kevin Zelaya}
\author[*]{Mohammad-Ali Miri}
\affil[ ]{Department of Electrical and Microelectronic Engineering, Rochester Institute of Technology, Rochester, New York 14623, USA}
\affil[*]{ammeme@rit.edu}

\begin{abstract}
Programmable photonics provides a versatile platform for optical information processing, but extending programmable transformations to large numbers of optical modes remains challenging. We investigate a modular architecture composed of unitary tiles arranged in alternating and staggered columns. Each physical tile interlaces fixed multiport mixers with programmable phase shifts, confining individual mixing operations to small groups of modes while connecting them across a larger processor. We analyze the relationship between circuit depth, intertile connectivity, and attainable transformations, distinguishing ideal programmable tiles from their physical phase-stack implementations. Numerical target-fitting results show a transition in the reconstruction errors at $R = 2K$, where $K$ and $R$ are the number of tile rows and columns, respectively. This design ultimately aims to enable large and reliable port-count optical photonic processors.
\end{abstract}

\setboolean{displaycopyright}{false} 

\begin{document}

\maketitle

\textbf{Introduction.} Matrix-vector multiplication lies at the core of unconventional analog computing paradigms, ranging from coherent optical neural networks and convolutional accelerators \cite{shen2017deep,bogaerts2020programmable} to programmable quantum transport simulators and photonic quantum processors \cite{harris2018linear}. Implementing an arbitrary $N_{\mathrm{total}}\times N_{\mathrm{total}}$ unitary operator $U \in \mathrm{U}(N_{\mathrm{total}})$ on a photonic chip requires independent control over $\dim_{\mathbb{R}}\mathrm{U}(N_{\mathrm{total}}) = N_{\mathrm{total}}^{2}$ real degrees of freedom, up to a global phase.

Typically, universal programmable photonics has relied on two primary paradigms. The first employs planar meshes of local $2\times 2$ Mach-Zehnder interferometers (MZIs) arranged in triangular Reck \cite{reck1994experimental} or rectangular Clements \cite{clements2016optimal} topologies, which admit analytical two-mode Givens rotations. However, MZI meshes require $N_{\mathrm{total}}(N_{\mathrm{total}}-1)/2$ directional couplers and an optical depth of $N_{\mathrm{total}}$ MZI stages; as $N_{\mathrm{total}}$ scales, cumulative insertion losses and fabrication errors in calibrated 50:50 beam splitters severely degrade unitary fidelity \cite{saygin2020robust,markowitz2023auto}. The second paradigm factorizes a unitary operator by interlacing fixed passive multiport mixing operators $F \in \mathrm{U}(N)$, realized via multimode interference (MMI), discrete Fourier transform (DFT) couplers, or evanescently coupled waveguide arrays, with diagonal programmable phase-shifter layers \cite{saygin2020robust,tanomura2022scalable,markowitz2023auto,zelaya2024goldilocks,alvarez2026universality,lopezpastor2021arbitrary,girouard2026near,markowitz2025photonic}. If the passive mixer $F$ satisfies a sufficient modal coupling density criterion (the ``Goldilocks'' regime) \cite{zelaya2024goldilocks,alvarez2026universality}, an $N$-port interlaced cascade achieves full universality with high compactness and intrinsic resilience to static fabrication disorder \cite{saygin2020robust,markowitz2023auto}. Nevertheless, scaling a single coupled-waveguide array to large port counts ($N_{\mathrm{total}} \gg 10$) increases the mixing coupling length required for edge-to-edge waveguide mixing and renders the global non-convex optimization landscape progressively harder to navigate \cite{zelaya2024goldilocks}. To overcome the scaling bottlenecks of both $2\times 2$ MZI meshes and multiport arrays, modular and fractal unitary decompositions have been explored across spatial and internal optical modes \cite{dhand2015realization,su2019hybrid,kumar2020optimal,basani2023self,goubault2026designs}.

\begin{figure}[!t]
  \centering
  \includegraphics[width=0.90\linewidth]{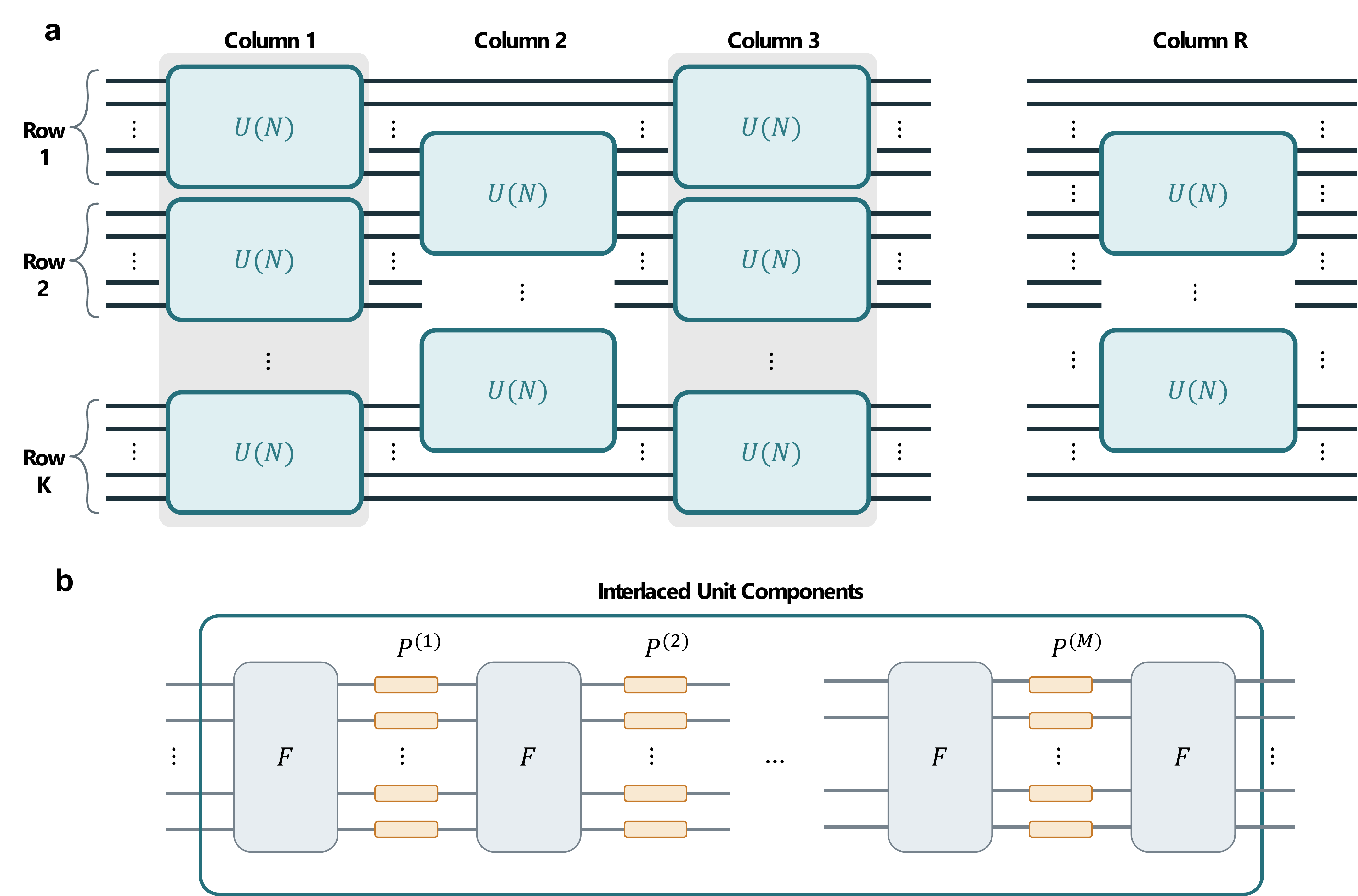}
  \caption{
  Proposed $KN$-port modular architecture. \textbf{a}~Brick-wall mesh of $K$ rows of even-$N$-port tiles $U(N)$ across $R$ columns, with even columns shifted by $N/2$ ports. \textbf{b}~Each $N\times N$ unit $U(N)$ interlaces $M+1$ passive mixers $F$ with $M$ phase-shifter layers $P^{(1)},\dots,P^{(M)}$.
  }
  \label{fig:arch}
\end{figure}

In this letter, we propose a brick-wall photonic architecture that employs discrete even-$N$-port unitary tiles ($N = 2n$) to construct modular, high-dimensional programmable processors (Fig.~\ref{fig:arch}\textbf{a}). By partitioning $N_{\mathrm{total}} = KN$ waveguides into $K$ modular rows of $N$-port interlaced units $U(N)$ and staggering even columns vertically by $N/2$ ports across $R$ sequential columns, our architecture generalizes the Clements topology \cite{clements2016optimal} from two-mode MZI Givens rotations to $(N/2)$-dimensional subspace block eliminations \cite{kumar2020optimal,basani2023self,sutton2009computing}. This design confines evanescent multiport coupling to compact local $N$-port blocks while achieving $\mathrm{U}(KN)$ universality at a critical column depth of $R_{\mathrm{crit}} = 2K$.

\textbf{Results.} The fundamental building block of the proposed brick-wall architecture, depicted in Fig.~\ref{fig:arch}\textbf{b}, is an even-$N$-port interlaced photonic unit composed of $M$ functional interlacing stages, configured with $M+1$ passive $N$-port waveguide-array layers $F$ and $M$ active programmable phase-shifter layers $P^{(m)}$ \cite{markowitz2023auto,zelaya2024goldilocks}. Within each $N\times N$ unit, a coherent optical field envelope $\mathbf{u} = [u_1, u_2, \dots, u_N]^\top \in \mathbb{C}^N$ transforms across the passive multiport propagation $F \in \mathrm{U}(N)$ and active diagonal phase modulation $P^{(m)} = \exp(i D^{(m)})$, where
\begin{equation}
  D^{(m)} = \mathrm{diag}\!\left(\phi_1^{(m)}, \phi_2^{(m)}, \dots, \phi_N^{(m)}\right), \quad \phi_n^{(m)} \in (0, 2\pi],
  \label{eq:phase_diag}
\end{equation}
for port index $n \in \{1,\dots,N\}$ and layer index $m \in \{1,\dots,M\}$. The transfer matrix $U \in \mathrm{U}(N)$ of a single unit is
\begin{equation}
  U\equiv U(\Phi) = F\,P^{(M)} F\, \cdots F\,P^{(1)} F = F \prod_{m=1}^{M} \left(P^{(m)} F\right),
  \label{eq:UN_block}
\end{equation}
where $\Phi=\bigcup_{m=1}^{M}D^{(m)}$ is the set of active phases in the unit.

Physically, each passive mixer $F$ is implemented as an array of $N$ evanescently coupled single-mode waveguides, each with propagation length $\ell$ and a gap size $s_{n}$ between the $n$-th and $(n+1)$-th waveguides for $n\in\{1,\ldots,N-1\}$. The overall field envelope $\mathbf{u}$ evolves according to coupled-mode theory~\cite{CMT}. The universality of the architecture described in~\eqref{eq:UN_block} has been thoroughly evaluated using numerical methods~\cite{markowitz2025photonic}, including a density criterion~\cite{markowitz2025photonic,markowitz2023auto,zelaya2024goldilocks} and a Fourier absorption criterion~\cite{alvarez2026universality}. Because factoring out one phase of each layer leaves $M(N-1)+1$ independent real parameters, $M=N+1$ active layers ($N(N+1)$ physical phase shifters) establishes standalone tile universality over $\mathrm{U}(N)$ while providing robustness to passive fabrication defects or nonfunctional active elements~\cite{saygin2020robust,markowitz2023auto}.

\begin{figure}[!t]
  \centering
  \includegraphics[width=0.90\linewidth]{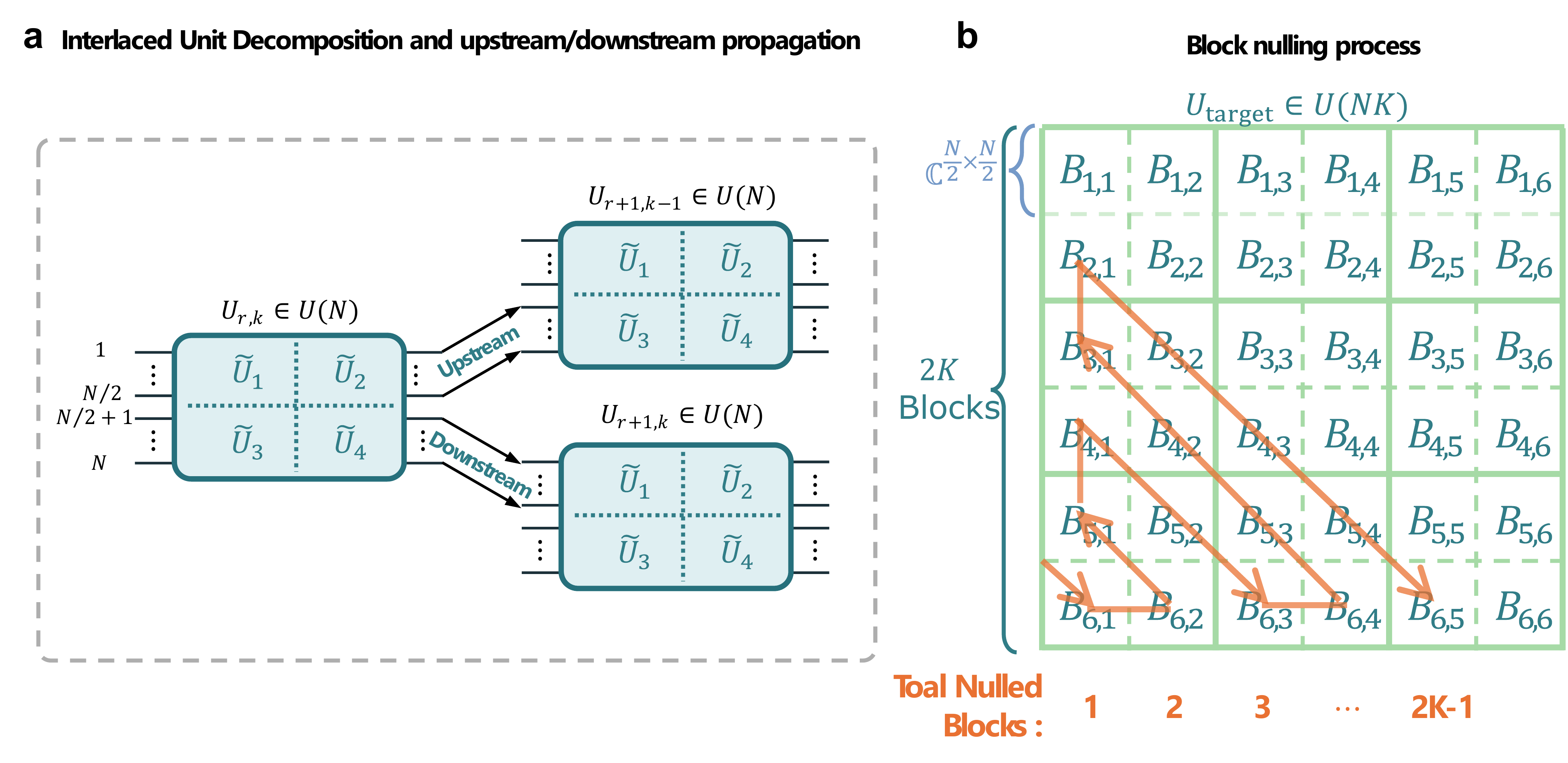}
  \caption{
  Subspace decomposition and nulling process. \textbf{a}~Partitioning of $U_{r,k} \in \mathrm{U}(N)$ into $(N/2)\times(N/2)$ sub-blocks $(\widetilde{U}_1,\dots,\widetilde{U}_4)$ and upstream/downstream routing into column $r+1$. \textbf{b}~Block grid of $U_{\mathrm{target}} \in \mathrm{U}(KN)$ ($K=3$) and alternating anti-diagonal nulling sweeps $s=1,\dots,2K-1$ (orange arrows).
  }
  \label{fig:decomp}
\end{figure}

\textit{Brick-wall modular $KN$-port architecture.} To scale the total port count to $N_{\mathrm{total}} = KN$ while keeping the evanescent mixing width fixed at $N$ waveguides, we embed the even-$N$-port tiles into a two-dimensional brick-wall mesh of $K$ spatial rows and $R$ sequential columns (Fig.~\ref{fig:arch}\textbf{a}). Let the $KN$ waveguides be indexed by $j \in \{1, 2, \dots, KN\}$ and grouped into $K$ contiguous rows of $N$ waveguides. Because $N$ is even, each $N$-port tile $U_{k,r}(N) \in \mathrm{U}(N)$ splits symmetrically into upper ($1, \dots, N/2$) and lower ($N/2+1, \dots, N$) sub-bundles of $N/2$ ports each, partitioning the tile into four square $(N/2)\times(N/2)$ sub-blocks $(\widetilde{U}_1, \widetilde{U}_2, \widetilde{U}_3, \widetilde{U}_4)$ (Fig.~\ref{fig:decomp}\textbf{a}). As illustrated in Fig.~\ref{fig:decomp}\textbf{a}, between adjacent columns the upper $(N/2)$-port output sub-bundle of tile $(k,r)$ routes upward along the ``upstream'' branch into the lower input half of tile $(k-1,r+1)$, whereas the lower $(N/2)$-port output sub-bundle routes downward along the ``downstream'' branch into the upper input half of tile $(k,r+1)$.

The modular circui alternates between two staggered column configurations across $r = 1, 2, \dots, R$. Each odd column ($r = 1, 3, \dots$) contains $K$ parallel tiles $U_{k,r}(N)$ ($k \in \{1,\dots,K\}$) aligned with the $K$ row partitions, coupling waveguides $(k-1)N+1$ through $kN$, resulting in the odd-column representation
\begin{equation}
  C_{\mathrm{odd}}^{(r)} = \bigoplus_{k=1}^{K} U_{k,r}(N) =
  \mathrm{diag}\!\left(U_{1,r}(N), \dots, U_{K,r}(N)\right) \in \mathrm{U}(KN).
  \label{eq:C_odd}
\end{equation}
In turn, each even column ($r = 2, 4, 6, \dots$) contains $K-1$ parallel tiles shifted vertically by $N/2$ ports to bridge adjacent rows; that is, the $k$-th tile $U_{k,r}(N)$ ($k \in \{1,\dots,K-1\}$) couples the lower $N/2$ ports of row $k$ with the upper $N/2$ ports of row $k+1$ (waveguides $(k-1)N + N/2 + 1$ through $kN + N/2$), while the outer $N/2$ boundary ports at the top of row $1$ and bottom of row $K$ traverse uncoupled identity waveguide sections $\mathbb{I}_{N/2}$ via
\begin{equation}
  C_{\mathrm{even}}^{(r)} = \mathbb{I}_{N/2} \oplus \left(\bigoplus_{k=1}^{K-1} U_{k,r}(N)\right) \oplus \mathbb{I}_{N/2} \in \mathrm{U}(KN).
  \label{eq:C_even}
\end{equation}
The global $KN\times KN$ transfer matrix $U_{\mathrm{total}} \in \mathrm{U}(KN)$ of an $R$-column processor is the ordered product
\begin{equation}
  U_{\mathrm{total}} = C^{(R)} C^{(R-1)} \cdots C^{(2)} C^{(1)}, \quad
  C^{(r)} = \begin{cases}
    C_{\mathrm{odd}}^{(r)}, & r \text{ odd}, \\
    C_{\mathrm{even}}^{(r)}, & r \text{ even}.
  \end{cases}
  \label{eq:U_total}
\end{equation}
This results in the total interlaced unit count
\begin{equation}
  N_{\mathrm{blocks}}(R) = K\lfloor \tfrac{R+1}{2} \rfloor + (K-1)\lfloor R/2 \rfloor.
  \label{eq:N_blocks_general}
\end{equation}

\textit{Subspace Block-QR Elimination and Critical Depth $R_{\mathrm{crit}} = 2K$.} To establish the critical column depth $R_{\mathrm{crit}}$ required to synthesize an arbitrary target $U_{\mathrm{target}} \in \mathrm{U}(KN)$, we partition the global Hilbert space $\mathcal{H} \cong \mathbb{C}^{KN}$ into $2K$ equal subspaces $\mathcal{H} = \bigoplus_{p=1}^{2K} \mathcal{H}_p$ of dimension $\dim\mathcal{H}_p = N/2$. Viewing $U_{\mathrm{target}}$ as a $2K\times 2K$ block matrix with square sub-blocks $B_{p,q} \in \mathbb{C}^{(N/2)\times(N/2)}$ ($p,q \in \{1,\dots,2K\}$; Fig.~\ref{fig:decomp}\textbf{b}), each tile $U_{k,r}(N)$ acts nontrivially on an adjacent subspace pair $\mathcal{H}_p \oplus \mathcal{H}_{p+1}$ ($p = 2k-1$ in odd columns; $p = 2k$ in even columns). Explicitly, we define the embedded single-tile unitary matrix $T_{\{p,p+1\}} \in \mathrm{U}(KN)$ for any $p \in \{1,\dots,2K-1\}$ as
\begin{equation}
  T_{\{p,p+1\}} \equiv \mathbb{I}_{(p-1)N/2} \oplus U_{k,r}(N) \oplus \mathbb{I}_{(2K-p-1)N/2} \in \mathrm{U}(KN),
  \label{eq:T_def}
\end{equation}
which acts as $U_{k,r}(N)$ on subspaces $\{p, p+1\}$ and as $\mathbb{I}_{N/2}$ on all remaining $2K-2$ subspaces, so that $C_{\mathrm{odd}}^{(r)} = \prod_{k=1}^{K} T_{\{2k-1,2k\}}$ and $C_{\mathrm{even}}^{(r)} = \prod_{k=1}^{K-1} T_{\{2k,2k+1\}}$.

For any stacked pair of square sub-blocks $(X^\top, Y^\top)^\top \in \mathbb{C}^{N\times(N/2)}$ with $X,Y \in \mathbb{C}^{(N/2)\times(N/2)}$, full QR decomposition $(X^\top, Y^\top)^\top = Q(R_{\mathrm{upper}}^\top, 0)^\top$ yields a unitary matrix $Q \in \mathrm{U}(N)$ and an upper-triangular matrix $R_{\mathrm{upper}} \in \mathbb{C}^{(N/2)\times(N/2)}$. By left-multiplying $(X^\top, Y^\top)^\top$ by a tile $U_{k,r}(N) \in \mathrm{U}(N)$ completely zeroes out the lower block $Y$,
\begin{equation}
  U_{k,r}(N)\begin{pmatrix} X \\ Y \end{pmatrix} = \begin{pmatrix} \tilde{X} \\ 0_{(N/2)\times(N/2)} \end{pmatrix},
  \label{eq:block_nulling}
\end{equation}
if and only if $U_{k,r}(N)$ takes the form
\begin{equation}
  U_{k,r}(N) = (L_1 \oplus L_2)\,Q^\dagger, \qquad L_1, L_2 \in \mathrm{U}(N/2),
  \label{eq:stabilizer_form}
\end{equation}
resulting in $\tilde{X} = L_1 R_{\mathrm{upper}}$, with $L_{1}$ and $L_{2}$ unconstrained unitary matrices. Thus, nulling a complex $(N/2)\times(N/2)$ sub-block imposes $2(N/2)^2 = N^2/2$ real constraints on $U_{k,r}(N)$ while leaving the block-diagonal output post-rotation $L_1 \oplus L_2 \in \mathrm{U}(N/2)\times\mathrm{U}(N/2)$ ($2(N/2)^2 = N^2/2$ real parameters) completely unconstrained. Dually, right-multiplying a horizontal pair $(X, Y) \in \mathbb{C}^{(N/2)\times N}$ by $U_{k,r}(N)^\dagger$ zeroes the left $(N/2)\times(N/2)$ block $X$ and updates $\tilde{Y} = R_{\mathrm{row}}^\dagger R_2^\dagger$ up to an arbitrary block-diagonal input pre-rotation $R_1 \oplus R_2 \in \mathrm{U}(N/2)\times\mathrm{U}(N/2)$.

\begin{figure*}[!t]
  \centering
  \includegraphics[width=0.74\linewidth]{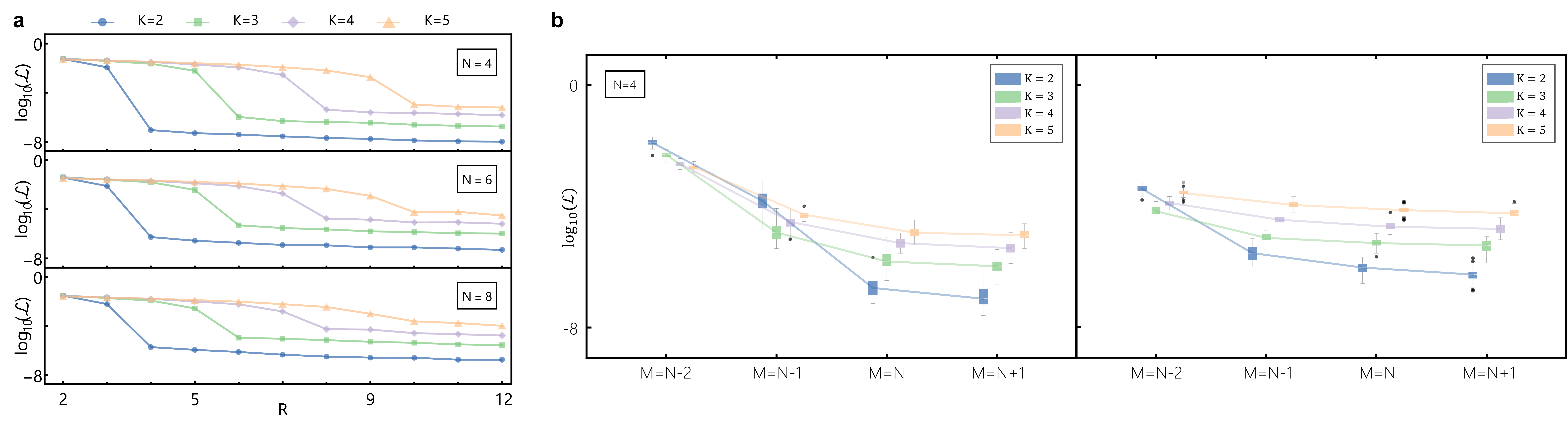}
  \caption{
  Numerical universality verification across $200$ Haar-random targets $U_{\mathrm{target}} \in \mathrm{U}(KN)$ for $K \in \{2,3,4,5\}$. \textbf{a}~Median log-scale Frobenius error $\log_{10}\mathcal{L}_{KN}$ versus column depth $R \in [2,12]$ for $N=4,6,8$, showing the $N$-independent cliff at $R_{\mathrm{crit}} = 2K$. \textbf{b}~Error distributions at $R = 2K$ versus local tile phase-layer count $M \in \{N-2, N-1, N, N+1\}$ for $N=4$ (left) and $N=6$ (right).
  }
  \label{fig:transition}
\end{figure*}

In~\eqref{eq:U_total}, using a single embedded tile~\eqref{eq:T_def} from the input end ($C^{(1)}$) right-multiplies $U_{\mathrm{target}}$ by $T_{\{q,q+1\}}^\dagger$ and mixes adjacent block-columns $(q,q+1)$, whereas using a tile from the output end ($C^{(R)}$) left-multiplies by $T_{\{p,p+1\}}^\dagger$ and mixes adjacent block-rows $(p,p+1)$. Using left-multiplication alone (row operations only, as in a Reck triangle \cite{reck1994experimental}) mixes non-zero entries in adjacent columns and overwrites previously created zeros unless delayed into a triangular mesh of depth $4K-3$. Instead, as shown by the orange trajectory in Fig.~\ref{fig:decomp}\textbf{b}, we eliminate all $\binom{2K}{2} = K(2K-1) = 2K^2 - K$ lower-triangular $(N/2)\times(N/2)$ blocks of $U_{\mathrm{target}}$ along successive anti-diagonals $s = 1, 2, \dots, 2K-1$ from the bottom-left block $B_{2K,1}$ by alternating between right-multiplications $R_s^\dagger = \prod_{j=1}^{s} T_{\{s-j+1,s-j+2\}}^{(s,j)\dagger}$ (column operations on odd sweeps $s$) and left-multiplications $L_s^\dagger = \prod_{j=s}^{1} T_{\{2K-s+j-1,2K-s+j\}}^{(s,j)\dagger}$ (row operations on even sweeps $s$) \cite{clements2016optimal,kumar2020optimal}
\begin{equation}
  \left(\prod_{s\text{ even}} L_s^\dagger\right) U_{\mathrm{target}} \left(\prod_{s\text{ odd}} R_s^\dagger\right) = D_{\mathrm{rem}} = \bigoplus_{p=1}^{2K} W_p,
  \label{eq:bidirectional_elim}
\end{equation}
where each sweep $s$ zeroes $s$ anti-diagonal blocks in parallel with existing zeros so that every linear combination of already-zeroed blocks remains zero. Once all $2K^2 - K$ lower-triangular blocks vanish, unitarity forces all upper-triangular blocks to vanish simultaneously, leaving the block-diagonal unitary $D_{\mathrm{rem}}$ with $W_p \in \mathrm{U}(N/2)$. Unwrapping the left and right factors in~\eqref{eq:bidirectional_elim} and absorbing $D_{\mathrm{rem}}$ into the terminal column packs all $2K^2 - K$ tiles into $K$ odd columns ($K$ tiles each) and $K$ even columns ($K-1$ tiles each), proving constructively that every $U_{\mathrm{target}} \in \mathrm{U}(KN)$ is realizable at $R_{\mathrm{crit}} = 2K$. This can be further verified by looking at the number of local tiles required to null all the lower-triangular blocks shown in Fig.~\ref{fig:decomp}\textbf{b}, which becomes $\sum_{k=1}^{2K-1}k=2K^2-K$. Using this number and the tile-count oin Eq.~\eqref{eq:N_blocks_general} results on $R=2K$.

\textit{Architecture Parameter Counting.} We derive the effective parameter dimension $D_{\mathrm{eff}}(R)$ of an $R$-column processor directly from the matrix product $U_{\mathrm{total}} = C^{(R)}\cdots C^{(1)}$. An $R$-column brick-wall mesh contains $R_{\mathrm{odd}} = \lceil R/2 \rceil$ odd columns ($K$ tiles each) and $R_{\mathrm{even}} = \lfloor R/2 \rfloor$ even columns ($K-1$ tiles each), giving a total tile count of $N_{\mathrm{blocks}}(R)$ in~\eqref{eq:N_blocks_general}. In $U_{\mathrm{total}} = C^{(R)}\cdots C^{(1)}$, the $K$ parallel tiles in the input column $C^{(1)} = \bigoplus_{k=1}^{K} U_{k,1}(N)$ have no tiles preceding them and contribute their full $K N^2$ real unitary parameters. Conversely, as depicted in Fig.~\ref{fig:decomp}\textbf{a}, every tile $U_{k,r}(N)$ among the remaining $N_{\mathrm{blocks}}(R) - K$ tiles in columns $r = 2, \dots, R$ receives its upper and lower $(N/2)$-port input sub-bundles along the ``downstream'' and ``upstream'' branches from column $r-1$. Factoring any tile in columns $r \ge 2$ at its input up to a block-diagonal unitary $G_1^\dagger \oplus G_2^\dagger \in \mathrm{U}(N/2)\times\mathrm{U}(N/2)$ transfers $(G_1 \oplus G_2)$ onto the output sub-blocks ($\widetilde{U}_3,\widetilde{U}_4$ and $\widetilde{U}_1,\widetilde{U}_2$) of the feeding tiles in column $r-1$ while leaving the product $C^{(r)}C^{(r-1)}$ unchanged, $U_{k,r}(N)(G_1^\dagger \oplus G_2^\dagger)(G_1 \oplus G_2) = U_{k,r}(N)$. Because $\dim_{\mathbb{R}}[\mathrm{U}(N/2)\times\mathrm{U}(N/2)] = 2(N/2)^2 = N^2/2$ input parameters per tile are absorbed into the already-counted parameters of the preceding column, each tile after $C^{(1)}$ contributes $N^2 - N^2/2 = N^2/2$ new real degrees of freedom. Summing columns $1,\dots,R$ gives the exact effective parameter dimension
\begin{equation}
  D_{\mathrm{eff}}(R) = K N^2 + \big[N_{\mathrm{blocks}}(R) - K\big]\frac{N^2}{2} = \frac{N_{\mathrm{blocks}}(R) + K}{2}\,N^2.
  \label{eq:D_eff}
\end{equation}
The evaluation of~\eqref{eq:D_eff} across the three depth regimes in Fig.~\ref{fig:transition} shows the sharpness in the universality transition
\begin{itemize}
  \setlength{\itemsep}{0.5pt}\setlength{\parskip}{0pt}
  \item \textbf{Under-determined ($R = 2K-1$, $N_{\mathrm{blocks}} = 2K^2 - 2K + 1$):} $D_{\mathrm{eff}}(2K-1) = [K^2 - \tfrac{1}{2}(K-1)]N^2 < K^2 N^2$ for all $K \ge 2$, leaving a strict subspace parameter deficit $\Delta D(2K-1) = K^2 N^2 - D_{\mathrm{eff}}(2K-1) = \tfrac{1}{2}(K-1)N^2$.
  \item \textbf{Critical universal ($R = 2K$, $N_{\mathrm{blocks}} = 2K^2 - K$):} Adding the $2K$-th column ($K-1$ tiles) supplies the exact missing $\tfrac{1}{2}(K-1)N^2$ cross-subspace parameters, yielding $D_{\mathrm{eff}}(2K) = K^2 N^2 = \dim_{\mathbb{R}}\mathrm{U}(KN)$ ($\Delta D(2K) = 0$).
  \item \textbf{Over-parameterized ($R = 2K+1$, $N_{\mathrm{blocks}} = 2K^2$):} \eqref{eq:D_eff} gives $(K^2 + \tfrac{1}{2}K)N^2 > K^2 N^2$, saturating $\mathrm{U}(KN)$ with $\tfrac{1}{2}K N^2$ surplus global degrees of freedom.
\end{itemize}

\textit{Local Parameter Deficiencies Effects.}
Equations~\eqref{eq:stabilizer_form} and~\eqref{eq:D_eff} show that individual $U(N)$ tiles need not be locally universal ($D_{\mathrm{tile}} = N^2 - \delta$) to preserve global $\mathrm{U}(KN)$ universality, provided two bounds hold. (i)~The \textit{local elimination bound} $N^2 - \delta \ge N^2/2$ ($\delta \le N^2/2$), ensuring each tile intersects~\eqref{eq:stabilizer_form}, and (ii)~the \textit{global supply bound} $N_{\mathrm{blocks}}(R)(N^2 - \delta) \ge K^2 N^2$, compensating the $K\delta$ deficit in $C^{(1)}$ via reduced inter-column overlap ($N^2/2 - \delta$) in Columns $2,\dots,R$. Combining both gives $\delta_{\max}(K,N,R) = \min\!\big(N^2/2,\, [1 - K^2/N_{\mathrm{blocks}}(R)]N^2\big)$, which at $R_{\mathrm{crit}} = 2K$ ($N_{\mathrm{blocks}} = 2K^2 - K$) results in
\begin{equation}
  \delta_{\max}(K, N)\Big|_{R=2K} = \left(\frac{K - 1}{2K - 1}\right) N^2 < \frac{N^2}{2}.
  \label{eq:even_deficit_law}
\end{equation}
For omitted whole phase layers ($\delta \approx cN$, reducing $M$ from $N+1$ to $N+1-c$), \eqref{eq:even_deficit_law} satisfies $\frac{K-1}{2K-1}N^2 \ge \frac{1}{3}N^2 \ge N$ for all even $N \ge 4$ and $K \ge 2$, guaranteeing that every tile at $R=2K$ can omit at least one full phase layer ($M = N$ \cite{markowitz2025photonic}) without losing global universality.

\textbf{Numerical Universality Verification.} To verify our results, we minimize the Frobenius reconstruction error norm
\begin{equation}
  \mathcal{L}_{KN}(\boldsymbol{\phi}) = \frac{1}{(KN)^2}\left\|U_{\mathrm{total}}(\boldsymbol{\phi}) - U_{\mathrm{target}}(KN)\right\|_F^2,
  \label{eq:frobenius_global}
\end{equation}
by optimizing all $N_{\mathrm{phases}}(R) = N_{\mathrm{blocks}}(R) N M$ phase shifters simultaneously via L-BFGS-B \cite{zelaya2024goldilocks,markowitz2025photonic} with closed-form analytical trace gradients. Figure~\ref{fig:transition}\textbf{a} plots the median converged error $\log_{10}\mathcal{L}_{KN}$ across $200$ Haar-random targets $U_{\mathrm{target}} \in \mathrm{U}(KN)$ versus column depth $R \in \{2,\dots,12\}$ for $K \in \{2,3,4,5\}$ and $N \in \{4,6,8\}$ (up to $KN = 40$, $\dim_{\mathbb{R}}\mathrm{U}(40) = 1600$).

\begin{table}[!t]
  \centering
  \caption{Architectural Scaling Comparison for a Universal $N_{\mathrm{total}} = KN$ Photonic Processor with Even-$N$ Tiles}
  \label{tab:scaling}
  \resizebox{\linewidth}{!}{%
    \begin{tabular}{@{}lccc@{}}
      \toprule
      \textbf{Metric} & \textbf{Clements MZI} \cite{clements2016optimal} & \textbf{Interlaced} \cite{zelaya2024goldilocks} & \textbf{Modular Brick-Wall (This Work)} \\
      \midrule
      Building block & $2\times 2$ MZI & $KN\times KN$ array & Even $N\times N$ tile $U\in\mathrm{U}(N)$ \\
      Critical column depth $R_{\mathrm{crit}}$ & $KN$ & $1$ tile ($KN+1$ layers) & $2K$ columns \\
      Total tiles $N_{\mathrm{blocks}}(R_{\mathrm{crit}})$ & $KN(KN-1)/2$ & $1$ & $2K^2 - K$ \\
      Active phase layers / tile $M$ & $2$ & $KN + 1$ & $N + 1 - c$ ($c \ge 1$ for $N \ge 4$) \\
      Max per-tile deficit $\delta_{\max}$ ($R_{\mathrm{crit}}$) & $0$ & $0$ & $\tfrac{K-1}{2K-1}N^2$ ($\ge N$ for $N\ge 4, K\ge 2$) \\
      Evanescent coupling width & $2$ modes & $KN$ modes & $N$ modes ($N \ll KN$) \\
      Mixer length per layer $\ell$ & $\ell_{\mathrm{DC}}$ & $\propto KN/\kappa$ & $\propto N/\kappa$ \\
      Splitter defect resilience & Low (50:50 BS) & High (dense $F$) & High (local dense $F$) \\
      \bottomrule
    \end{tabular}%
  }
\end{table}

Consistent with~\eqref{eq:D_eff}, Fig.~\ref{fig:transition}\textbf{a} exhibits three regimes: (i)~for $R \le 2K-1$, $\log_{10}\mathcal{L}_{KN}$ decreases gradually to a pre-transition shoulder ($\approx -2.0$ to $-2.8$) at $R = 2K-1$; (ii)~adding one column to reach $R_{\mathrm{crit}} = 2K$ ($R = 4, 6, 8, 10$) triggers an $N$-independent cliff of $2.5$--$5$ orders of magnitude ($\log_{10}\mathcal{L}_{KN} < -4$, reaching $-7.1$, $-6.3$, and $-5.7$ at $K=2$); and (iii)~for $R > 2K$, $\log_{10}\mathcal{L}_{KN}$ settles onto a convergence floor between $\sim 10^{-8}$ ($N=4, K=2$) and $\sim 10^{-4}$ ($N=8, K=5$, up to $3960$ phases).

To test the deficiency law~\eqref{eq:even_deficit_law} directly, Fig.~\ref{fig:transition}\textbf{b} sweeps the local phase-layer count $M \in \{N-2, N-1, N, N+1\}$ at $R = 2K$ for $N=4$ and $N=6$. Removing one mask per tile increased the median reconstruction loss by approximately $0.07–0.35$ across the eight tested configurations ($N = 4, 6 and K = 2–5$). In turn, reducing $M = N-1$ degrades the median error to values from approximately $-4.0$ to $-4.8$ (most sharply at $K=2$, where the deficit approaches the $R=2K$ ceiling), and $M = N-2$ makes all configurations underdetermined.

Table~\ref{tab:scaling} summarizes these scaling advantages over $2\times 2$ Clements MZI meshes \cite{clements2016optimal} and $KN$-port interlaced arrays \cite{zelaya2024goldilocks}: our modular topology restricts evanescent coupling to $N$ waveguides per tile ($\ell_{\mathrm{mod}} \propto N/\kappa$) while tolerating per-tile parameter deficiencies up to $\delta_{\max} = \frac{K-1}{2K-1}N^2$ at $R = 2K$. Table~\ref{tab:scaling} compares our modular block-interlaced topology against canonical $2\times 2$ Clements MZI meshes \cite{clements2016optimal} and $KN$-port interlaced arrays \cite{zelaya2024goldilocks}. In cases where a $KN$-port waveguide array requires coupling across all $KN$ channels in every passive layer ($\ell_{\mathrm{mono}} \propto KN/\kappa$), our modular architecture restricts evanescent coupling to $N$ channels per tile ($\ell_{\mathrm{mod}} \propto N/\kappa$) while tolerating a per-tile parameter deficiency up to $\delta_{\max} = \frac{K-1}{2K-1}N^2$ at $R = 2K$.

\textbf{Conclusion.} We have analyzed a modular brick-wall photonic architecture that achieves universal $\mathrm{U}(KN)$ transformations across $KN$ optical modes by staggering even-$N$-port interlaced unitary tiles by $N/2$ waveguides across $R$ columns. Using subspace block-QR elimination, column-by-column parameter counting, and L-BFGS-B numerical optimization across $N \in \{4,6,8\}$ and $K \in \{2,3,4,5\}$, we proved and verified a sharp universality transition at the critical column depth $R_{\mathrm{crit}} = 2K$, which persists even when individual tiles operate with a parameter deficiency up to $\delta_{\max} = \frac{K-1}{2K-1}N^2$ (allowing at least one omitted $N$-phase layer per tile for all even $N \ge 4$ and $K \ge 2$). This hierarchical framework decouples local multiport wave mixing from global port scaling, offering a compact, disorder-resilient platform for large-scale programmable photonics.

\begin{backmatter}
\bmsection{Funding} This project is supported by the U.S. Air Force Office of Scientific Research (AFOSR) Award \# FA9550-25-1-0200, the National Science Foundation (NSF) Award \# 2329021, and the National Aeronautics and Space Administration (NASA) Award \# 80NSSC26K1252.
\smallskip
\bmsection{Disclosures} The authors declare no conflicts of interest.
\smallskip
\bmsection{Data availability} Data underlying the results presented in this paper may be obtained from the authors upon reasonable request.
\smallskip
\end{backmatter}

\bibliography{references}

\end{document}